\documentclass[10pt,
							 twocolumn,
							 superscriptaddress,
							 english,
							 prl,
							 showpacs,
							 floatfix,
							 aps
							]{revtex4-1}

\usepackage{siunitx}
\usepackage{ulem}
\usepackage{tabularx}
\usepackage{xfrac}
\usepackage{graphicx}
\usepackage{color}
\usepackage{amsthm}
\usepackage{amsmath}
\usepackage{amssymb}
\usepackage{esint}
\usepackage{multirow}
\usepackage{bigdelim}

\usepackage{upgreek}
	\usepackage{amsmath}
	\usepackage{makeidx}
	\usepackage{amsfonts}
	\usepackage[usenames,dvipsnames]{pstricks}
	\usepackage{subfigure}

\usepackage[utf8]{inputenc}
\usepackage{xspace}

\makeatletter

\usepackage{soul}
\usepackage[backref=none,
bookmarksnumbered=true,
bookmarks=true,
bookmarksopen=true,
colorlinks=true,
citecolor=blue,
linkcolor=blue,
anchorcolor=green,
urlcolor=blue,unicode=false]{hyperref}

\newcommand{\didv}{\ensuremath{{\rm d}I/{\rm d}V}\xspace}
\newcommand{\diidvv}{\ensuremath{{\rm d}^2I/{\rm d}V^2}\xspace}
\newcommand{\Ef}{\ensuremath{E_\mathrm{F}}\xspace}

\makeatother

\usepackage{babel}

\begin{document}
\title{Correlation of Vibrational Excitations and Electronic Structure\\ with Submolecular Resolution}

\author{Daniela Rolf}
\affiliation{Fachbereich Physik, Freie Universit\"at Berlin, 14195 Berlin, Germany}
\author{Friedrich Maaß}
\affiliation{Physikalisch-Chemisches Institut, Universit\"at Heidelberg, 69120 Heidelberg, Germany}
\author{Christian Lotze}
\affiliation{Fachbereich Physik, Freie Universit\"at Berlin, 14195 Berlin, Germany}
\email{c.lotze@fu-berlin.de}
\author{ Constantin\ Czekelius}
\affiliation{Institut f\"ur Organische Chemie und Makromolekulare Chemie,
 Heinrich-Heine-Universit\"at D\"usseldorf, 40225 D\"usseldorf, Germany}
\author{Benjamin W. Heinrich}
\affiliation{Fachbereich Physik, Freie Universit\"at Berlin, 14195 Berlin, Germany}
\author{Petra Tegeder}
\affiliation{Physikalisch-Chemisches Institut, Universit\"at Heidelberg, 69120 Heidelberg, Germany}
\author{Katharina J. Franke}
\affiliation{Fachbereich Physik, Freie Universit\"at Berlin, 14195 Berlin, Germany}





\begin{abstract}
The detection of vibrational excitations of individual molecules on surfaces by scanning tunneling spectroscopy does not obey strict selection rules but rather propensity rules. The experimental verification of these is challenging because it requires the independent variation of specific parameters, such as the electronic structure, while keeping the vibrational modes the same.  Here, we make use of the versatile self-assembled structures of Fe-tetra-pyridyl-porphyrin molecules on a Au(111) surface. These exhibit different energy-level alignments of the frontier molecular orbitals, thus allowing the correlation of electronic structure and detection of vibrations. We identify up to seven vibrational modes in the tunneling spectra of the molecules in some of the arrangements, whereas we observe none in other structures.  We find that the presence of vibrational excitations and their distribution along the molecule correlates with the observation of energetically low-lying molecular states. This correlation allows to explain the different numbers of vibrational signatures for molecules embedded within different structures as well as the bias asymmetry of the vibrational intensities within an individual molecule. Our observations are in agreement with a resonant enhancement of vibrations by the virtual excitation of electronic states.

\end{abstract}
\maketitle

\section{Introduction}
Since the seminal discovery of molecular excitations in single molecules on surfaces by inelastic electron tunneling spectroscopy (IETS) in a scanning tunneling microscope (STM) \cite{stipe_single_molecule_1998}, various studies reported on the observation of vibrational excitations of chemisorbed as well as physisorbed molecules \cite{ho_single_molecule_2002, komeda_chemical_2005, lauhon_single-molecule_2000,pascual_adsorbate-substrate_2001,kim_single-molecule_2002, heinrich_cascade_2004, okabayashi_2008, franke_excitation_2010, Meierott_2016}. For some molecules, a large set of vibrational excitations was observed, whereas others did not show a single vibrational mode. Moreover, the resolution of vibrational excitations depended on the specific substrate underneath the molecules \cite{franke_excitation_2010}. While selection rules exist that predict the intensity of vibrational modes for optical spectroscopic techniques, such as Raman and infrared spectroscopy, there are no such universal criteria in scanning tunneling spectroscopy (STS). Instead, a set of propensity rules has been proposed \cite{lorente_symmetry_2001,troisi_molecular_2006,paulsson_unified_2008}.

Theoretical models suggested different mechanisms leading to vibrational excitations within the neutral molecule, \textit{i.e.}, at energies below a molecular ion resonance. The common requirement is that the energy of the tunneling electron exceeds the excitation energy of the vibration \cite{jaklevic_1966}.
The first mechanism is an excitation via impact scattering of the tunneling electron, which is independent of existing molecular resonances \cite{lorente_theory_2000,lorente_mode_2004}. The corresponding change of the differential conductance (\didv) is positive but typically very small \cite{lorente_theory_2000}, such that excitations by impact scattering are rarely detected in spectroscopy.
The second mechanism depends on the electronic structure. Even for the case that the electron energy falls below a molecular resonance, virtual excitations to low-lying molecular states enhance the excitation cross section of vibrations \cite{persson_inelastic_1987,lauhon_single-molecule_2000,pascual_adsorbate-substrate_2001,kim_single-molecule_2002,lorente_mode_2004, katharina_j_franke_and_jose_ignacio_pascual_effects_2012}.
Note, that this regime is distinctly different from the resonant regime, in which the vibrational excitation occurs in the charged state, such that vibronic peaks would appear on top of the molecular resonance in the \didv spectra \cite{liu_vibronic_2004,frederiksen_dynamic_2008}. Due to the limited lifetime of the vibronic states in molecules on metal substrates, the vibronic peaks would just contribute to an overall broadening of the electronic resonances.
In the resonant regime, the observation of vibronic peaks has been improved by an increase of the lifetime of the excited states by either  ultrathin insulating layers \cite{qiu_vibronic_2004, Krane2018}, layers of organic molecules \cite{katharina_j_franke_and_jose_ignacio_pascual_effects_2012} or bulky molecular groups attached to the molecule, which were employed to decouple the molecule from the substrate \cite{matino_electronic_2011}.

Off-resonance inelastic vibrational excitations are typically not observed when the molecules lie on decoupling layers, because the probability of virtual excitations is suppressed due to the up-shift of the molecular ion resonances. It is therefore difficult to achieve optimum conditions for the observation of strong inelastic signals. One notable example of achieving these conditions was reported by Ohta \textit{et al.} on Fe-phthalocyanine bilayers on Ag(111) \cite{ohta_enhancement_2013}. The authors also noted a peculiar difference in the observation of different vibrational modes in different adsorption configurations, which could not be explained within the currently available models and underlines the quest for a microscopic understanding of propensity rules. However, propensity rules also include the electronic structure of the molecule and the tip \cite{lorente_symmetry_2001,garcia_lekue_simulation_2011,pavlicek_symmetry_2013, Meierott_2016} and the dipole moment of the modes \cite{garcia_lekue_simulation_2011}.
The experimental verification of propensity rules is challenging, because the variation of one of these parameters is typically achieved by exchanging the molecule, which often entails a variation in several properties.
To avoid these complexities, previous studies compared the IETS signal of the same molecule on different surfaces \cite{franke_excitation_2010}, with different orientations \cite{Meierott_2016,meierott_line_2017}, or at different adsorption sites \cite{gawronski_physisorption_2014}.
Despite all those studies, there is still no comprehensive picture to date that allows to predict the intensity of molecular vibrations in IETS. Therefore, a more detailed understanding is necessary.

\begin{figure}[t]
\includegraphics[width=0.5\textwidth]{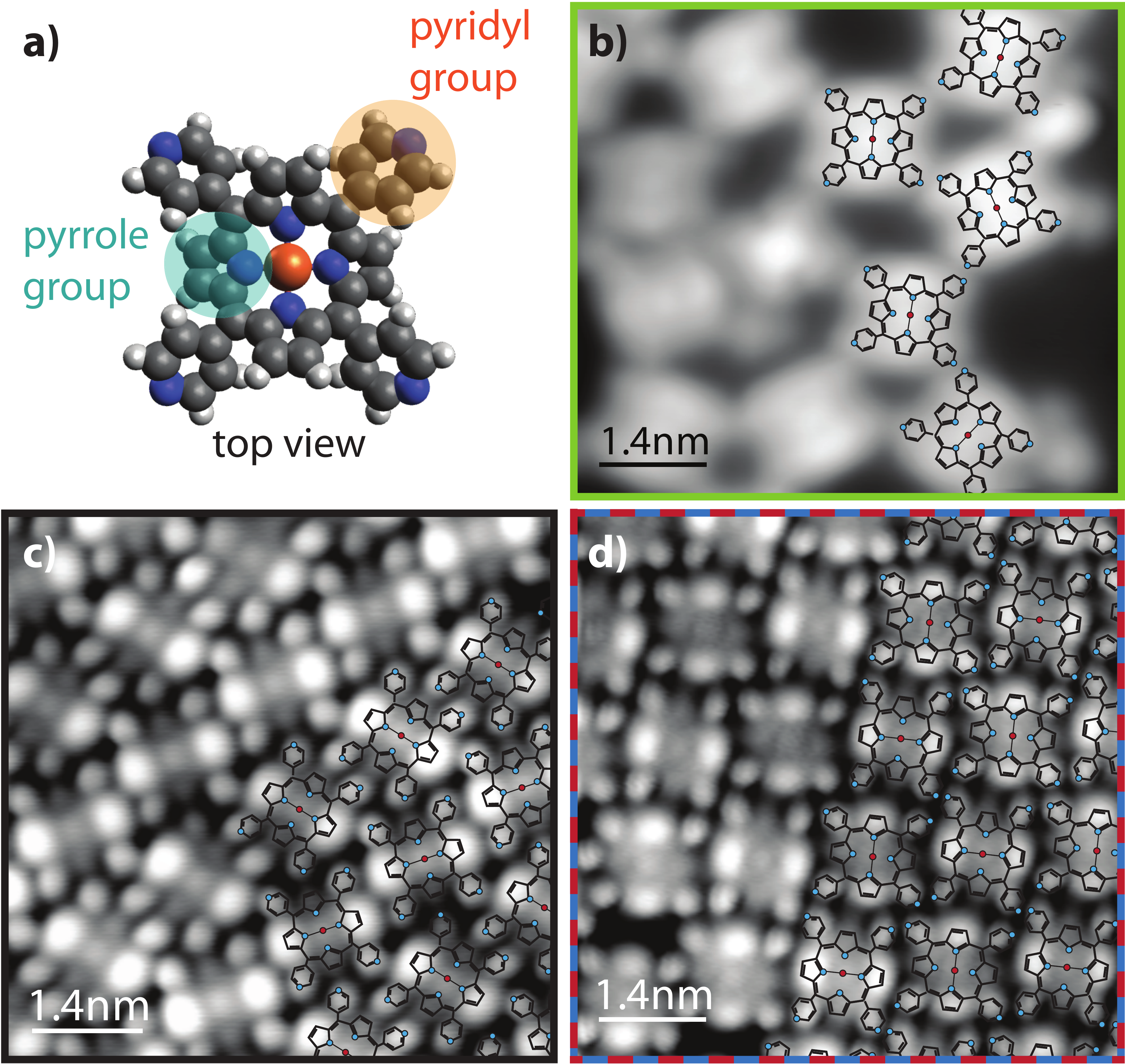}
\caption{a) Schematic structure of the FeTPyP molecule, with the pyrrole groups marked in blue and the freely rotatable pyridyl groups indicated in orange. Due to its flexibility, the molecules adapt a saddle-shape configuration upon adsorption. b-d) Different arrangements of FeTPyP on Au(111): b) Disordered structure of FeTPyP; c) Densely-packed structure of FeTPyP; d) Staggered arrangement of FeTPyP with an alternating orientation of the molecules. In this arrangement, two types of molecules can be identified by their spectral properties. In all STM topographies, the saddle-shape deformation of the molecules is apparent, as two pyrrole groups of the molecules appear higher than the other two. The color of the boxes around the images indicates the colors of the spectra in Fig.\,\ref{Fig2}). Topographies recorded at \SI{440}{\mV}, \SI{93}{\pA} (b), \SI{200}{\mV}, \SI{200}{\pA} (c) and \SI{230}{\mV}, \SI{160}{\pA} (d).}
\label{Fig1}
\end{figure}

Our approach is to use a molecule that exhibits different conformations on a surface. Thereby, the electronic structure undergoes some variations while local vibrational modes, such as C--H- and C--C-stretching and bending modes are not expected to change much. Such a system allows for a direct correlation of electronic structure and vibrational excitation intensity.
We use the flexible molecule Fe-5,10,15,20-tetra-4-pyridyl-porphyrin (FeTPyP) (shown in Fig.\,\ref{Fig1}a) as a model system to investigate the role of the electronic structure for the detection of vibrations in tunneling spectroscopy within the same species. The FeTPyP molecules consist of an Fe center embedded in a porphyrin core with four pyridyl moieties \cite{Fetpyp,Fetpyp2}. This molecule adapts different structures in molecular assemblies, concomitant with a change in the energy-level alignments. Using STS, we find up to nine inelastic steps in some of the molecular structures while others are featureless. We identify seven of these steps as molecular vibrations whereas the other two correspond to spin excitations. We show that the intensity of vibrational excitations can be correlated to the energy of the frontier molecular orbitals in the different structures. We explicitly show that an asymmetric electronic structure around the Fermi level leads to an asymmetry in the vibrational excitation intensity when comparing peaks at positive and negative bias voltage.

\section{Methods}
All experiments were performed under ultra-high vacuum conditions with \textit{in-situ} sample preparation in different chambers. The clean Au(111) samples were prepared by subsequent cycles of sputtering and annealing. The FeTPyP-Cl molecules were deposited from a Knudsen cell evaporator at \SI{410}{\celsius} onto  a Au(111) sample held at room temperature. During the deposition, the molecules are dechlorinated, such that the Fe-center changes its oxidation state from +3 to +2  \cite{heinrich_change_2013,Ben_FeOEP_2015, gopakumar_transfer_2012, Rolf_Visualizing_2018}.
The STM measurements were performed in two different low-temperature STMs, working at \SI{1.1}{K} and \SI{4.5}{K}, as indicated in the respective figure captions. The images were taken in constant-current mode, whereas \didv  spectra were recorded in open feedback-loop conditions with a lock-in amplifier. HREELS measurements were performed at \SI{90}{K} at an incident electron energy of \SI{3.5}{\eV}. DFT calculations were performed for single molecules in gas phase using the Gaussian 09 package \cite{Gaussian} and employing the B3PW91 hybrid functional. We used the 6-31g* basis set for C, N and H atoms and the LanL2dz basis set with effective core potentials (ECP) for Fe.

\section{Results and Discussion}
Using scanning tunneling microscopy, we observe three different structures of FeTPyP on Au(111).
An STM topography of the first observed molecular arrangement at sub-monolayer coverage is shown in Fig.\,\ref{Fig1}b. It reveals a disordered arrangement of the FeTPyP molecules with the pyridyl moieties of neighboring molecules facing each other. We speculate that atoms that were unintendedly co-deposited during the evaporation act as bonding nodes to the electron-rich pyridyl endgroups. These appear flatter than typically observed in pure molecular structures. The rather flat pyridyl groups lead to an enhancement of the screening and hybridization of the molecule with the substrate's electronic states \cite{xianwen_chen_conformational_2017}. This is reflected in low-lying molecular states (see below).

In a different preparation, we find two densely-packed arrangements without any adatoms. In the first, the molecules are aligned in two alternating rows of parallel molecules (Fig.\,\ref{Fig1}c). In the second structure (Fig.\,\ref{Fig1}d), the molecules are ordered in a staggered arrangement, with their saddle orientation rotated by \SI{90}{\degree} with respect to the neighboring molecules. This structure contains two different types of FeTPyP molecules, which exhibit distinct spectroscopic properties.
The saddle deformation, which is observable in all structures, is a result of molecule--substrate interactions \cite{auwarter_controlled_2007}. These force the pyridyl groups from the perpendicular orientation with respect to the porpyhrin plane into an inclined orientation. As a consequence of steric hindrance between adjacent hydrogen atoms in the pyrrole and pyridyl groups, two pyrrole groups bend up while the other pair is bent down.

\begin{figure*}[t]
\includegraphics[width=13cm]{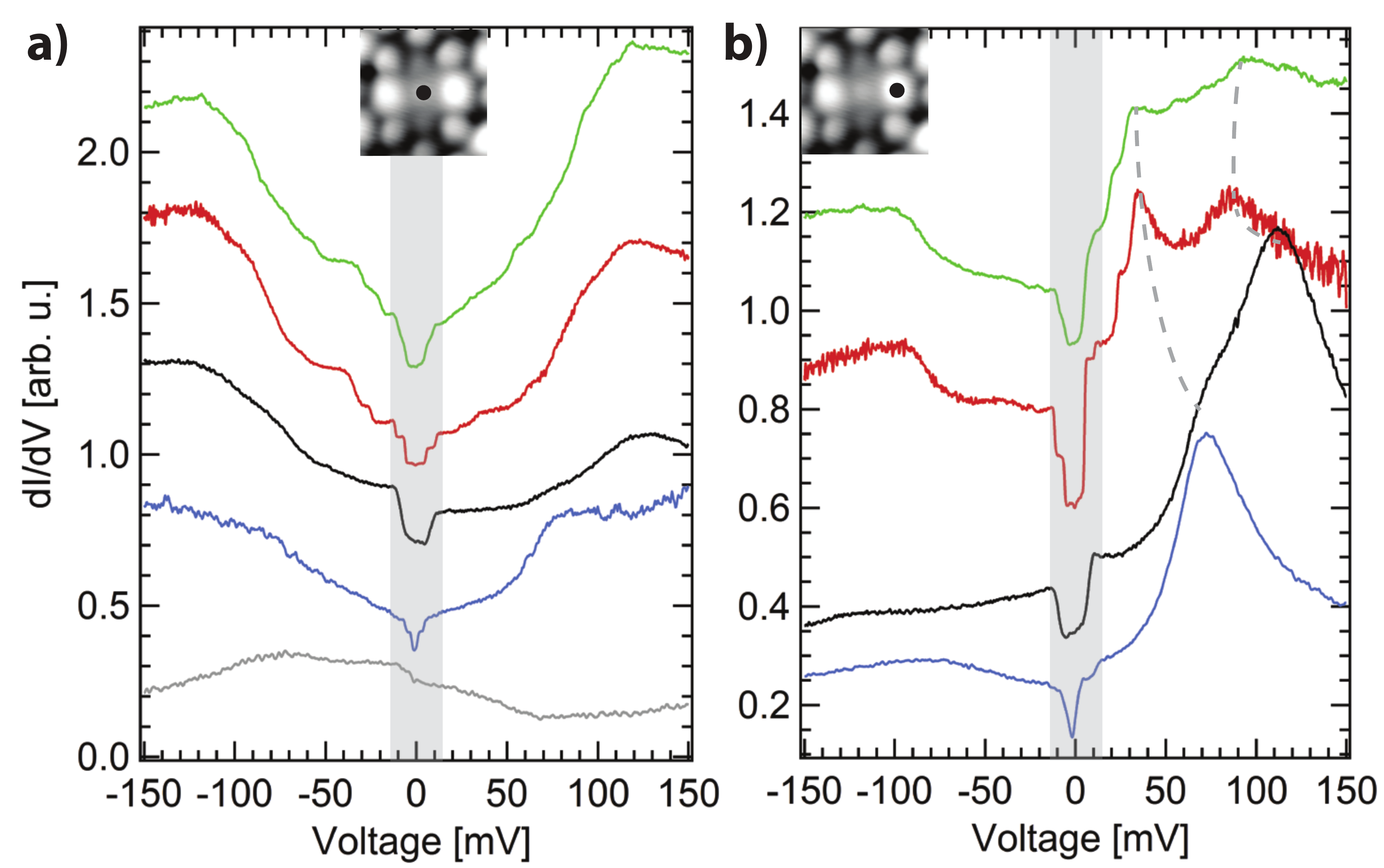}
\caption{Comparison of the vibrational signatures a) in the center and b) on the ligand of FeTPyP molecules  in the different arrangements (compare to color of the boxes in Fig.\,\ref{Fig1}). Overall, all FeTPyP molecules show bias-symmetric lineshapes in the center, and more asymmetric lineshapes on the ligand. The featureless gray spectrum corresponds to a non-metallated TPyP molecule. Different numbers of steps can be observed in the spectra. The gray-shaded area indicates those steps that originate from spin excitations. Feedback opened at green: \SI{150}{\mV}, \SI{3}{\nA} with V$_\text{mod}=\SI{1}{\mV}$, $T=\SI{4.5}{\K}$; red and blue: \SI{200}{\mV}, \SI{2}{\nA} with V$_\text{mod}=\SI{0.5}{\mV}$, $B=\SI{0.5}{\tesla}$, $T=\SI{1.1}{\K}$; black and gray: \SI{200}{\mV}, \SI{2}{\nA} with V$_\text{mod}=\SI{1}{\mV}$, $T=\SI{4.5}{\K}$.}
\label{Fig2}
\end{figure*}

To compare the properties of the FeTPyP molecules in the different arrangements, \didv spectra in the energy range of $\pm\SI{150}{\meV}$ are shown in Fig.\,\ref{Fig2}, which were recorded both in the center and on the upper pyrrole group of the molecules. The color of the spectra corresponds to the color of the boxes around the respective molecular structures in Fig.\,\ref{Fig1}. For comparison, a spectrum of one of the metal-free TPyP molecules, which are occasionally found on the surface, is shown as well. It neither  exhibits peaks nor steps. In the center of all Fe-containing molecules (Fig.\,\ref{Fig2}a), an almost bias-symmetric shape on the scale of $\pm\SI{150}{\meV}$ can be observed. However, upon closer inspection there are some differences. The red and green spectra exhibit a strong increase in conductance at $\pm\SI{80}{\meV}$. At similar energies, the conductance increase in the black spectrum seems broader. The blue spectrum is almost featureless at negative bias voltages and exhibits a shoulder at about $\SI{70}{\meV}$ at positive bias voltage.
Additionally, all of these spectra show one or two steps at low bias voltages (see grey shaded area), which are attributed to inelastic spin excitations \cite{Rolf_Visualizing_2018}.
Moreover, the green and red curve show a variety of steps at higher energies, whose position and relative intensities are independent of the employed Au tip. These higher-lying steps are assigned to vibrational excitations. 

In Fig.\,\ref{Fig2}b, we probe the presence/absence of steps on the upper pyrrole groups of the molecules. While the low-energy spin excitations are present in all molecules also on the ligand, the overall lineshapes of the \didv spectra differ quite drastically between the molecules in the different structures. The blue and the black spectra show a rather flat \didv curve at negative bias voltages, whereas the green and red spectra exhibit a broad shoulder at around \SI{-84}{\mV}. Moreover,   only the red and green spectra exhibit some additional steps at positive bias voltages. These steps are followed by double-peak structures at about \SI{35}{\mV} and \SI{90}{\mV}. In the black spectrum these peaks appear shifted to higher energies, \textit{i.e.}, to \SI{70}{\mV} and \SI{115}{\mV} (see grey dashed line as guide to the eye of the shift). The blue spectrum exhibits a single peak at \SI{72}{\mV}.

\begin{figure}[t]
\includegraphics[width=0.5\textwidth]{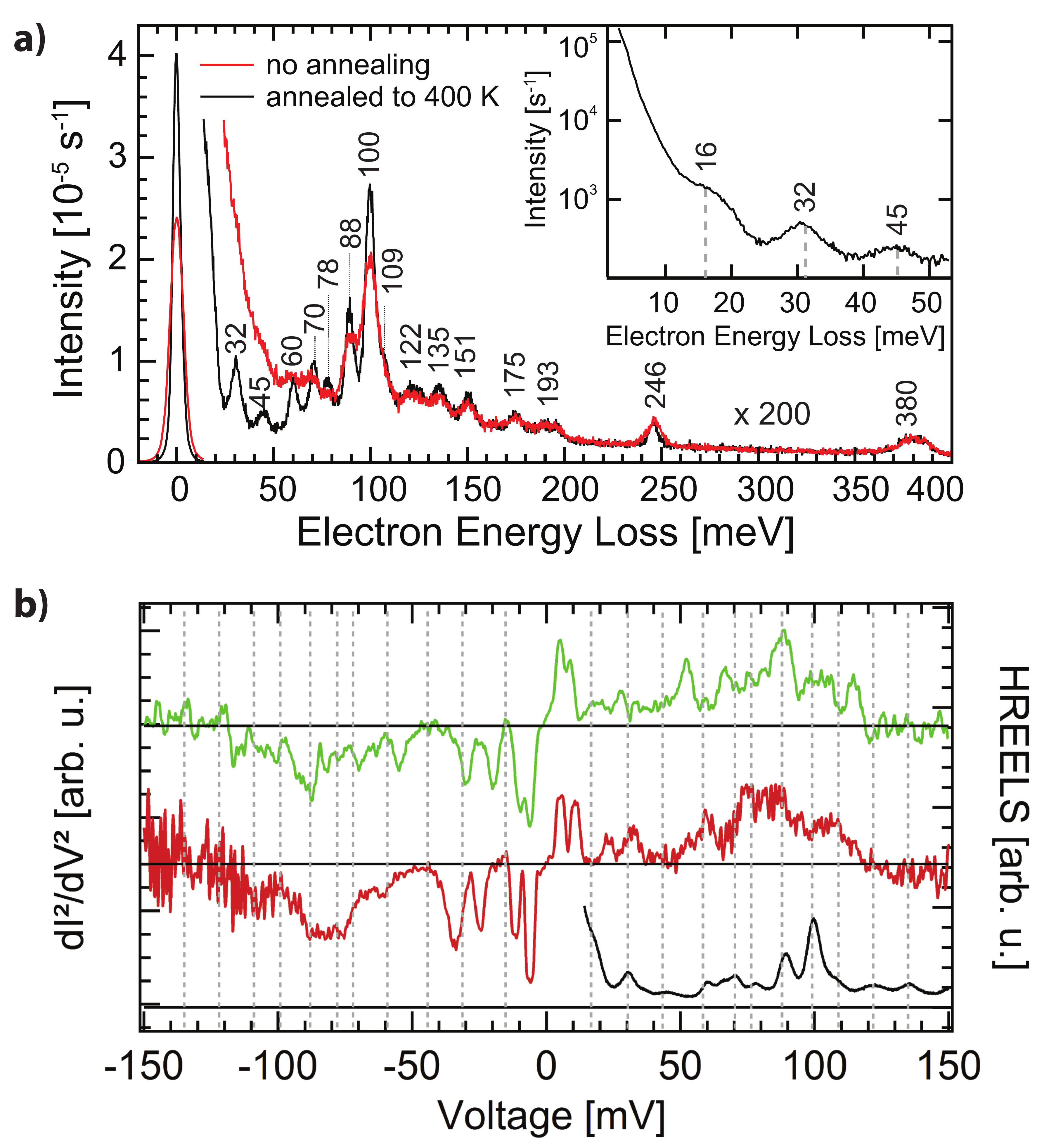}
\caption{Identification of molecular vibrations: a) HREEL spectrum recorded with a primary electron energy of \SI{3.5}{\eV} in specular scattering geometry before annealing (red) and after annealing to 400 K (black). The energy resolution measured as FWHM of the elastically
scattered electrons (elastic peak) is \SI{4.7}{\meV} (black) and \SI{7.9}{\meV} (red). The inset shows the low-energy region of the annealed sample, confirming the absence of vibrational modes below \SI{16}{\meV}. b) Comparison between the numerically derived \diidvv spectra (green and red) corresponding to the \didv spectra of Fig.\,\ref{Fig2}a and the HREEL spectrum of the annealed sample (black). The dashed lines indicate the positions of the vibrational modes as determined from the HREEL spectrum. The two lowest-lying peaks in the \diidvv spectra, are assigned to spin excitations.}
\label{Fig3}
\end{figure}

To confirm the assignment of the higher-lying steps as vibrational excitations, we performed HREELS measurements.
HREELS is a complementary method to identify molecular vibrations by inelastic electron scattering \cite{Ibach_1982, Maass_2016}. The incident electron energy (here \SI{3.5}{\eV}) exceeds the energies for vibrational transitions such that  vibrational excitations lead to energy losses  of the scattered electrons. Spectra taken on a monolayer deposited at room temperature and measured in specular scattering geometry (thus containing dipole and impact scattered electrons) are shown in Fig.\,\ref{Fig3}a (red).  Apart from the elastic peak, several vibrations are visible. Post-annealing to \SI{400}{\kelvin} results in an intensity-rise of the elastic peak and better resolved vibrational peaks (black  curve in Fig.\,\ref{Fig3}a) while the vibrational energies  are unchanged. The larger elastic peak intensity  as well
as the higher energy resolution (around \SI{4.7}{\meV} FWHM for
the black and \SI{7.9}{\meV} for the red spectrum) suggests a
higher degree of long-range order of the molecular layer after annealing.

However, we note that the existence of vibrational excitations is independent of the preparation conditions. Indeed, we repeated the measurements on several preparations without a change of the vibrational energies. A close-up view on the low-energy loss region (see inset of Fig.\,\ref{Fig3}a) indicates that the lowest vibrational mode lies at \SI{16}{\meV}. The absence of peaks at even lower energies corroborates the interpretation of the lowest two steps in the \didv spectra as spin excitations.

For better comparison of vibrational energies detected by HREELS and IETS, we plot the \diidvv spectra of the molecules in the green and red structure (Fig.\,\ref{Fig3}b). The steps in the \didv spectra correspond to peaks and dips in the \diidvv  signals at positive and negative bias voltage, respectively. The dashed lines indicate the peak positions of the vibrational modes as deduced from the HREEL spectrum. Besides the two lowest-energy peaks in the \diidvv signal, which have been discussed above, we find similarities between the vibrational modes from the IETS and from the HREELS measurements.

 \begin{table*}[t]
\centering
\caption{Assignment of the peaks in the HREEL spectra to steps in the \didv signal of the staggered and the disordered arrangements of FeTPyP, together with a general assignment of the modes by comparison to DFT calculations (b3pw91/genecp). All energies are given in meV.  }
\small
\begin{tabularx}{\textwidth}{llllp{2em}}
\hline
\textbf{HREELS} & \textbf{IETS (stagg.)} & \textbf{IETS (disord.)} &\textbf{Mode} \\
\hline
16 & 24.4 & 21& \rdelim\}{2}{*}[buckling/breathing  modes, Fe-tapping] \\
32 & 33.9 &30.5 \\
45 & - & -& \rdelim\}{10}{*}[C-C/C-H stretching and bending modes]  \\
60 & 61.0& 54 & \\
70 & - & 69 & \\
78 & 80.4& - &   \\
88 & - & 90 &   \\
100 & - & - &   \\
109 & 107& 104 &\\
122 & - &115 &  \\
135 & - & -&  &  \\
151 & - & -&  &  \\
\hline
\end{tabularx}
\label{tab1}
\end{table*}

A more detailed comparison of the vibrational energies is compiled in Tab.\,\ref{tab1}. We note that there are differences in the mode energies of a few meV between the HREELS data as well as between the staggered and disordered structure probed by IETS. We suggest that the different mode energies in IETS can be qualitatively understood by the different molecular conformations. The deviation from the HREELS data is roughly of the same size. However, HREELS is an ensemble method, such that an average energy both of several molecules and of different structures  is determined, whereas \didv spectroscopy determines the vibrational energies of a specific molecule. Moreover, the excitation mechanism of the two spectroscopic methods is different, which might account for further discrepancies in the spectra.

To gain qualitative insights into the origin of the modes and of the deviations, we performed DFT calculations (see Methods)
of isolated FeTPyP molecules with two different dihedral angles ($\ang{25}$ and $\ang{60}$), which mimic different degrees of the saddle-shape distortion. Indeed, the energies of the modes of these molecules differ. However, it is difficult to draw a one-to-one correspondence between these modes and the experimentally observed ones. The calculations reveal more than 100 different modes in the considered energy window. Therefore, we only categorize the modes at low and high energies. The low-energy modes correspond to symmetric and asymmetric buckling and stretching modes of the pyrrole groups, as well as Fe tapping modes. Since these modes involve a deformation of the pyrrole groups and the Fe--N distance within the molecule, their energies are expected to depend on the exact conformation of the molecule on the surface.
At higher energies, the vibrational modes constitute in-plane and out-of-plane stretching and bending modes of the C--H and C--C bonds. Similar vibrational modes at comparable energies were also observed on double-layer FePc molecules on Ag(111) \cite{ohta_enhancement_2013}, in agreement with similar intramolecular bonds in these molecules.

In Fig.\,\ref{Fig2} we saw that the vibrational excitations were only resolved in the \didv spectra of some of the molecular structures and with strong bias asymmetries. To understand these variations, we first consider possible effects of the tunneling barrier and the tip's electronic structure.
Small asymmetries of the signal intensities at opposite bias voltages might be explained by an asymmetry of the tunneling barrier \cite{Lauhon_effects_2001}. However, barrier-induced differences in the signal intensity should occur similarly at all positions across the molecule, which is in contrast to our experimental findings of almost symmetric intensities in the center.
Moreover, the symmetry and electronic structure of the tip was shown to affect the intensity of the vibrational fingerprints \cite{pavlicek_symmetry_2013,garcia_lekue_simulation_2011}. However, as the measurements in the center of the molecule and on the ligand were all performed using the same tip, this cannot explain the different inelastic signals.

Instead, by comparison to the position of close-lying resonances in the \didv spectra, we find that the intensity of the vibrational modes across the molecules follows the localization of the molecular electronic states.
The overall symmetric shape of the \didv spectra in the center of the molecules (Fig.\,\ref{Fig2}a) suggests the presence of occupied and unoccupied molecular states symmetrically around the Fermi level.
In contrast, the energies of molecular states on the ligand  are asymmetric with bias polarity, as the spectra in Fig.\,\ref{Fig2}b only reveal states at positive bias voltages.
The absence of features in the \didv spectra around \Ef on the metal-free porphyrin (gray curve in Fig.\,\ref{Fig2}a) indicates that the conductance in the FeTPyP molecules mainly stems from Fe d states.  This interpretation is supported by recent studies on Fe-tetra-phenyl-porphyrin (FeTPP) on Au(111)  \cite{carmen_rubio_verdu_orbital_selective_2017}, which showed similar lineshapes.
In that case, the density of states (DoS) around the Fermi level in the center of the molecule was assigned to the half-occupied d$_{z^2}$  orbital, which mostly interacts with the substrate. The resonance at positive bias voltages on the upper pyrrole groups was assigned to an empty hybrid state between the Fe d$_{yz}$ orbital and ligand states.
Given the structural and spectroscopic similarities between FeTPyP and FeTPP, this assignment is probably also valid for our case.

This electronic structure allows us to correlate the electronic structure with the vibrational sensitivity.
We first analyze the observations in the center of the molecules. In the absence of any molecular states close to \Ef, the excitation cross section of molecular vibrations is negligible. This is the case for the metal-free molecule. In contrast, the red and green spectrum show the largest increase of conductance at $\pm\SI{80}{\meV}$. The energetic proximity of occupied and unoccupied d states thus enhances the excitation cross section of the molecular vibrations at both bias polarities. A broadening of these states as in the black and blue spectrum largely suppresses the inelastic excitation probability.
On the ligand, the situation becomes even more interesting. In this case, there are only unoccupied states close to \Ef, which stem from a different orbital than in the center. The presence of this state enhances the vibrational excitations for the red and green molecule at positive bias voltages. This hybrid state is shifted away from \Ef in the blue and black spectra, such that the IETS intensity is below our experimental resolution. At negative bias voltages, none of the molecular species exhibits occupied states sufficiently close to \Ef.

All observations thus agree with the picture of an enhancement of the IETS signal by the presence of a low-energy molecular resonance. The lower the energy of the virtual excitation of this resonance, the more probable is this excitation. Therefore, also the cross section of the inelastic excitations increases.

\section{Conclusions}

We resolved several molecular vibrations of FeTPyP on a Au(111) substrate by a combination of HREEL and \didv spectroscopy.
A comparison of different structures of FeTPyP on a Au(111) substrate showed a varying number of inelastic steps between \SI{20}{\meV} and \SI{120}{\meV} in the corresponding \didv spectra, which were associated with excitations of molecular vibrations.
Interestingly, the intensity of the inelastic steps varied between the molecular structures and within the individual molecules. The simultaneous resolution of molecular orbital energies suggests a correlation of the vibration intensity with the proximity of the molecular orbital to the Fermi level. This correlation immediately reflects the applicability of the resonant-enhancement model \cite{persson_inelastic_1987}, which describes the vibrational excitation via virtual excitations of molecular resonances.

Funding by the ERC Consolidator grant "NanoSpin" (K. J. F.) and by the International Max Planck Research School "Functional Interfaces in Physics and Chemistry" (D. R.) is gratefully acknowledged.
F. M. and P. T. acknowledge funding by the German Research Foundation (DFG) through collaborative research center SFB1249 "N-Heteropolycylces
as Functional Materials" (project B06).



%

\end{document}